\documentstyle[prd,floats,aps]{revtex}
\input psfig
\begin{document}
\title{The general solution of the quantum Einstein equations?}

\author{Rodolfo Gambini\\{\em Instituto de F\'{\i}sica, Facultad de
Ciencias,\\Tristan Narvaja 1674, Montevideo, Uruguay}}

\author{Jorge Pullin\\  {\em Center for Gravitational Physics and Geometry}\\
{\em Department of Physics, 104 Davey Lab,}\\ {\em The Pennsylvania State 
University,}\\
{\em University Park, PA 16802}}

\maketitle
\begin{abstract}
We suggest how to interpret the action of the quantum Hamiltonian
constraint of general relativity in the loop representation as a skein
relation on the space of knots. Therefore, by considering knot
polynomials that are compatible with that skein relation, one
guarantees that all the quantum Einstein equations are solved.  We
give a particular example of such invariant and discuss the
consistency of the constraint algebra in this approach.
\end{abstract}

\vspace{-8cm} 
\begin{flushright}
\baselineskip=15pt
CGPG-96/3-1  \\
gr-qc/9603019\\
\end{flushright}
\vspace{7cm}

The canonical approach provides a setting to address the
nonperturbative aspects of the quantization of gravity. Within this
approach, the introduction of the Ashtekar new variables \cite{As86}
has allowed to formulate the theory in terms of a connection, as
opposed to the usual approach in which the metric plays the
fundamental role. Having the theory cast in terms of a connection 
allows to consider traces of holonomies (Wilson
loops) as fundamental variables and to build a representation (the
loop representation) in which wavefunctions are functions of loops.

Because general relativity is invariant under diffeomorphisms, in the
loop representation one deals with functions of loops that have to be
invariant under deformations of the loop. Therefore the states of
quantum gravity have to be knot invariants. This was the key insight
that allowed Rovelli and Smolin in 1988 \cite{RoSm88} to show that
knot theory was crucially related to quantum gravity.

This connection is only ``kinematical'', in the sense that {\em any}
theory of a connection that is invariant under diffeomorphisms is also
related to knot theory. No input is needed from the particular
Lagrangian or equations of motion of Einstein's theory to establish
the connection.  This may raise the following question: up to what
point are the structures and ideas of knot theory, which never took
into account Einstein's equations, geared up to this newly discovered
connection with gravity?

The point of this essay is to suggest that a growing amount of
evidence has accumulated over the last years that indicates that an
unsuspected connection may exist between knot theory and the detailed
structure of the quantum Einstein equations. The main idea can be
summarized in that the action of the Hamiltonian constraint of quantum
gravity, when formulated in terms of knot space, is a skein
relation. Skein relations are the defining relations for knot
polynomials. Therefore it is natural that the solutions of the
Hamiltonian constraint should be related to knot polynomials. We will
show that a very nontrivial solution that was found through lengthy
direct calculations a few years ago, is actually compatible with the
skein relation induced by the Hamiltonian and therefore provides a
concrete detailed example of the idea we are putting forward.

The departing point is the action of the Hamiltonian constraint of quantum
gravity in the loop representation formulated in the lattice that we
have recently introduced \cite{GaPu96}. The action can be described
very simply through the following picture. It is non-zero only at
points where there is a triple intersection in the loops, and it acts
by converting the triple intersection in a pair of double
intersections, one for each pair of strands entering the intersection,
\begin{equation}
\hat{H} \psi(\eta_i) = \sum_{\rm pairs}\left[ \psi(\eta_1)-\psi(\eta_2)\right]
\label{eq1}
\end{equation}
where the action on one typical pair is depicted in figure 1. For
non-straight through intersections the action is different, we will
not discuss that case here for reasons of space, but all results
claimed for the straight through case go through for the other cases
as well.

Although this Hamiltonian was first proposed
in the context of a fully regularized lattice framework, it is
immediate to take over its action to the continuum theory.
\begin{figure}
\hskip 1cm \psfig{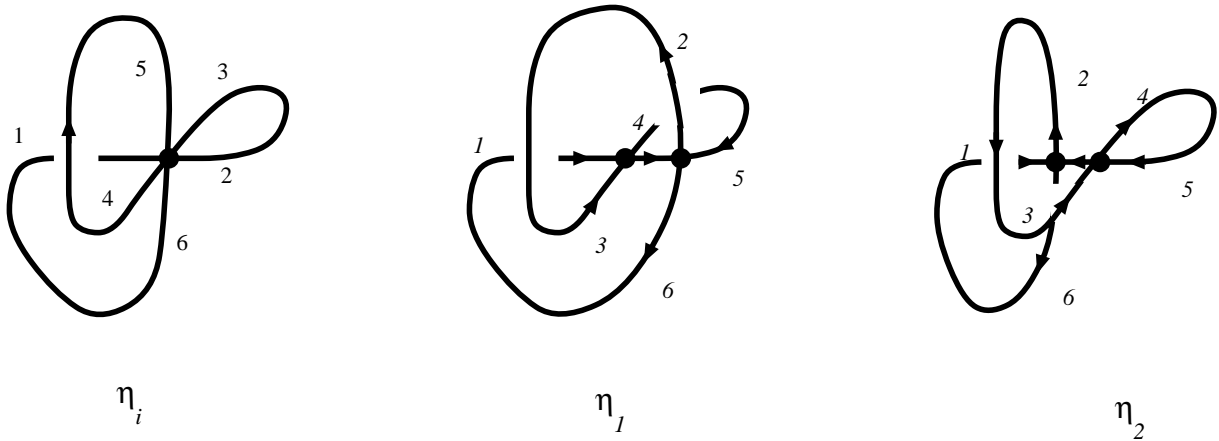}
\caption{The action of the Hamiltonian deforms the argument of the
wavefunction from $\eta_I$ to $\eta_\pm$.}
\label{fig1}
\end{figure}
Solutions to the constraint in knot space are invariants such that
when evaluated on knots like $\eta_1$ and $\eta_2$ give contributions
such that the sum in the right hand side of equation (\ref{eq1})
cancels.  Let us construct an explicit example of such an
invariant. It was shown some time ago \cite{BrGaPuprl} that the second
coefficient of the Conway \cite{Co} polynomial was a solution of the
Hamiltonian constraint of quantum gravity in the loop
representation. This result was first established in a formal way in
the continuum. Later it was shown in a regularized context via the
extended loop representation, albeit via the introduction of a
delicate counterterm in the Hamiltonian \cite{DiBaGaGr}. We will here
show that it is actually very easy to see that it satisfies condition
(1) and therefore it is a solution. In order to see this we need to
consider the skein relations that define the second coefficient of the
Conway polynomial. These can be found and suitably 
 generalized to intersecting loops using the recent results
of reference \cite{GaPu96b} by looking at the second coefficient
of the expansion in the variable $x$ of the Jones polynomial evaluated
for $q=e^x$. The result is,
\begin{eqnarray}
C_2(\raisebox{-3pt}{\psfig{figure=l+.eps,height=4mm}}) -
C_2(\raisebox{-3pt}{\psfig{figure=l-.eps,height=4mm}}) &=&
C_1(\raisebox{-3pt}{\psfig{figure=l0.eps,height=4mm}})\\
C_2(\raisebox{-3pt}{\psfig{figure=li.eps,height=4mm}}) &=&
{1\over 2} (C_2(\raisebox{-3pt}{\psfig{figure=l+.eps,height=4mm}}) +
C_2(\raisebox{-3pt}{\psfig{figure=l-.eps,height=4mm}})) +{1\over 8} 
(C_0(\raisebox{-3pt}{\psfig{figure=l+.eps,height=4mm}}) +
C_0(\raisebox{-3pt}{\psfig{figure=l-.eps,height=4mm}}))\\
C_2(\raisebox{-3pt}{\psfig{figure=lv.eps,height=4mm}}) &=&
\label{eqlv}
C_2(\raisebox{-3pt}{\psfig{figure=l0.eps,height=4mm}})\\
C_2(\raisebox{-3pt}{\psfig{figure=lo.eps,height=4mm}}) &=& 0,
\end{eqnarray}
where $C_0=2^{n_c}/2$ where $n_c$ is the number of connected
components of the knot ($C_0=1$ for a single component knot ) and $C_1$
is an invariant that coincides with the Gauss linking number if the
involved loop has two components and zero otherwise. The Gauss linking
number $lk(\gamma_1,\gamma_2)$ is a very simple invariant that given
two curves $\gamma_1,\gamma_2$ measures how many times one of them
pierces a surface that has the other curve as a boundary. It behaves
in an ``Abelian'' manner with respect to composition of curves
$lk(\gamma_1\circ\gamma_2,\gamma_3)=lk(\gamma_1,\gamma_3)+
lk(\gamma_2\circ\gamma_3)$ and also with respect to retracings of
curves, $lk(\gamma,\eta^{-1})=-lk(\gamma,\eta)$.

The above relations imply that for the second coefficient we can turn
an intersection into an upper or double crossing ``at the price'' of a
term proportional to the linking number and another one to the 
number of connected components,
\begin{equation}
C_2(\raisebox{-3pt}{\psfig{figure=li.eps,height=4mm}}) = 
C_2(\raisebox{-3pt}{\psfig{figure=l-.eps,height=4mm}}) +
{1\over 2} C_1(\raisebox{-3pt}{\psfig{figure=l0.eps,height=4mm}}) 
+{1\over 4}C_0(\raisebox{-3pt}{\psfig{figure=l-.eps,height=4mm}}) =
C_2(\raisebox{-3pt}{\psfig{figure=l+.eps,height=4mm}}) -
{1\over 2} C_1(\raisebox{-3pt}{\psfig{figure=l0.eps,height=4mm}})
+{1\over 4}C_0(\raisebox{-3pt}{\psfig{figure=l+.eps,height=4mm}}). 
\label{price}
\end{equation}

We can then eliminate the double intersection to the left of $\eta_1$
and the one to the right in $\eta_2$ using (\ref{price}) and the
intersection to the right of $\eta_1$ and the one to the left of
$\eta_2$ using (\ref{eqlv}) in the action of the Hamiltonian (1) in
such a way as to be left with $C_2$ evaluated in two topologically
equivalent knots. Therefore their contributions cancel. The price for
lifting and lowering the lines is a collection of terms involving the
linking number, that cancel each other by virtue of the Abelian nature
of $C_1$ and $C_0$ \cite{GaPu96}.  This proof is remarkably simpler
than the original derivation in the continuum, which had to be tackled
with computer algebra \cite{BrGaPuprl}, or even the extended loop
version which had several delicate issues of regularization involved
\cite{DiBaGaGr}.

What we accomplished here is to isolate the topological action of the
Hamiltonian constraint. The Hamiltonian constraint is an operator that
is not diffeomorphism invariant, it depends on a point. Therefore one
cannot expect to have a realization of it in knot space as an
operator.  The best one can hope for is to define its kernel. What we
have done to capture this concept is to implement the kernel of the
Hamiltonian as a skein relation in knot space. This has an added
bonus: one does not need to be concerned about the problem of the
constraint algebra. Operators have to satisfy algebras, skein
relations do not. An analogy of this in a more familiar context would
be given by considering the Gauss law. If one formulates the theory in
terms of a connection, the Gauss law is a quantum operator and one has
to worry about its constraint algebra. It one goes to the loop
representation, one is in the kernel of the Gauss law. The only
remnant of the Gauss law are the Mandelstam identities. These one
treats as relations and one does not worry about their algebra.

The relations we presented here come about because the regularized
action of the Hamiltonian constraint of quantum gravity in loop space
always can be written as a non diffeomorphism invariant pre-factor
that is point and regularization dependent times a function of loops
evaluated in a loop that is deformed from the original loop. By
separating the action into a ``topological'' and ``local'' part and
implementing the ``topological'' part as a skein relation what we are
doing is to construct the physical space of states of the theory.  In
this space one could represent diffeomorphism invariant observables
for the theory that one could use to perform physical measurements.

Viewing quantum gravity in the space of knots brings in a new light
the problem of the degrees of freedom of the theory. We are saying
that the physical states of gravity are knot invariants (compatible
with the Mandelstam identities) that satisfy the skein relation
induced by the Hamiltonian. This is not enough to characterize an
invariant. There are many invariants that satisfy this (an example was
$C_2$ as discussed above). The resulting freedom corresponds to the
degrees of freedom of quantum gravity. This suggests a new way of how
to handle the local degrees of freedom when one quantizes a field
theory as quantum general relativity that is topological and yet has
local degrees of freedom.

The above claims can only be considered as preliminary, but there are
a series of consistency checks that one can envisage carrying out in
the immediate future. On one hand, the ideas exposed do not depend on
the number of dimensions of spacetime. They could therefore be probed
carefully in the well-understood domain of gravity in $2+1$
dimensions.  In that case it is known that the Ashtekar version of the
constraints and the Witten one (that simply requires that the
curvature vanish) are equivalent only under a certain set of
assumptions. A detailed study should show which set of constraints
is the skein relation proposed here more closely related to. In the
Regge-Ponzano version of $2+1$ gravity it is known \cite{Rei} that the
Hamiltonian constraint leads to a relation among states that could be
viewed as a skein relation in two dimensions. The solution to it, in
the case of a trivial topology, is uniquely determined (as it should,
since there are no degrees of freedom in that case). This appears as a
direct counterpart in $2+1$ dimensions of the ideas we present here in
the $3+1$ context. In this latter context there are other checks that
can be carried out. For instance, it is known that the Kauffman
bracket knot polynomial should be a state of quantum gravity with a
cosmological constant \cite{BrGaPunpb} (it is the loop transform of
Kodama's Chern--Simons state). It is straightforward to write a skein
relation corresponding to the Hamiltonian constraint with a
cosmological constant (the cosmological constant adds only a rerouting
at triple intersections \cite{GaPuBa}). It would be remarkable if the
Kauffman bracket's skein relations are compatible with this
relation. Moreover, calculations in the extended loop representation
\cite{Gr} suggest that the third coefficient of the Jones polynomial
is not annihilated by the constraint but its action is proportional to
a topological rearrangement of the loops. Again one could check that
the skein relation presented here yields the same result.

All of this constitutes a mounting set of evidence that a deep
connection at a dynamical level exists between knot theory and quantum
gravity. The fact that notions of knot theory that were developed
independently of any theory of gravity are now being used to write in
a very detailed way the dynamics of quantum general relativity cannot
be understressed.

We wish to thank Abhay Ashtekar and Mike Reisenberger for discussions.
This work was supported in part by grants NSF-INT-9406269,
NSF-PHY-9423950, NSF-PHY-9396246, research funds of the Pennsylvania
State University, the Eberly Family research fund at PSU and PSU's
Office for Minority Faculty development. JP acknowledges support of
the Alfred P. Sloan foundation through a fellowship. We acknowledge
support of Conicyt and PEDECIBA (Uruguay).

\end{document}